\documentclass[aps,prb,twocolumn,superscriptaddress]{revtex4}
\newcommand{\bv}[1]{\ensuremath{\mathbf{#1}}}
\newcommand{\mrm}[1]{\ensuremath{\mathrm{#1}}}

\begin{document}

\title{Impact of long-range interactions on the disordered vortex 
lattice}

\author{J.A.G.~Koopmann}
\affiliation{Theoretische Physik, ETH-H\"onggerberg, CH-8093 
Z\"urich,
Switzerland}

\author{V.B.~Geshkenbein}
\affiliation{Theoretische Physik, ETH-H\"onggerberg, CH-8093 
Z\"urich,
Switzerland}
\affiliation{L.D. Landau Institute for Theoretical Physics, 117940
Moscow, Russia}

\author{G.~Blatter}
\affiliation{Theoretische Physik,
ETH-H\"onggerberg, CH-8093 Z\"urich, Switzerland}

\date{February 10, 2003}

\begin{abstract}
The interaction between the vortex lines in a type-II 
superconductor is
mediated by currents. In the absence of transverse screening this 
interaction
is long-ranged, stiffening up the vortex lattice as expressed by 
the dispersive
elastic moduli. The effect of disorder is strongly reduced, 
resulting in a
mean-squared displacement correlator $\langle u^2(\bv{R},L) 
\rangle \equiv
\langle [\bv{u(R},L)- \bv{u(0},0)]^2 \rangle$ characterized by a
mere logarithmic growth with distance. Finite screening cuts the 
interaction on
the scale of the London penetration depth $\lambda$ and limits 
the above
behavior to distances $R<\lambda$. Using a functional 
renormalization group
(RG) approach, we derive the flow equation for the disorder 
correlation
function and calculate the disorder-averaged mean-squared 
relative displacement
$\langle u^2(\bv{R}) \rangle  \propto\ln^{2\sigma} (R/a_0)$. The 
logarithmic
growth ($2\sigma=1$) in the perturbative regime at small 
distances [A.I.~Larkin
and Yu.N.~Ovchinnikov, J.\ Low Temp.\ Phys.\ \textbf{34}, 409 
(1979)] crosses
over to a sub-logarithmic growth with $2\sigma=0.348$ at large 
distances.
\end{abstract}
\maketitle

\section{Introduction}
Disordered elastic systems have attracted much attention in the 
last decade.
They have applications in various physical systems, such as 
charge density
waves,\cite{gruener1988} domain walls in 
Ising-ferromagnets,\cite{lemerle1998}
and vortices in type-II superconductors.\cite{blatter1994} The 
most interesting
physical properties of such disordered elastic systems are their 
structural
(static) order and their dynamical response under external 
forces. The dynamics
involves a finite critical force with creep
\cite{feigelman1989,fisher1991,chauve2000} at small- and depinning
\cite{narayan1993,nattermann1992} at large forces. Here, we 
concentrate on the
structural aspects as expressed through the displacement 
correlator $\langle
u^2(\bv{r}) \rangle \equiv \langle [\bv{u(r)-u(0)}]^2 \rangle$ 
(angular
brackets denote average over disorder). In general, the 
mean-squared
displacement grows with distance as $\langle u^2(\bv{r})\rangle 
\propto
|\bv{r}|^{2\zeta}$, where $\zeta$ is the wandering exponent 
characterizing the
roughness of the elastic manifold; for $\zeta>0$ the disorder is 
relevant and
the manifold is rough.\cite{larkin1970}

The generic model describing the competition between elasticity,
quenched disorder, and possibly thermal fluctuations involves an
isotropic homogeneous elastic medium (manifold) characterized by
$d$ internal dimensions ($\bv{r}$) and an $N$-component vector
field for the displacement $\bv{u(r)}$ ($[d+N]$-model). The system
we consider in the following is the flux-line lattice in a type-II
superconductor submitted to quenched disorder. The distortions of
the lattice are described by three internal dimensions
$\bv{r\equiv(R},L)$ and a two-component displacement field
$\bv{u(r)}$ ($[3+2]$-model).

A closer inspection shows that the underlying physical properties
of the vortex lattice infer interesting modifications of the
generic structural characteristics, both quantitatively and even
qualitatively (a similar type of specific properties is found in
other systems, e.g., in Wigner crystals). The underlying structure
of the vortex lattice introduces a number of additional length
scales in the problem: \emph{i})~On a microscopic level, the
flux-line lattice is built from interacting lines; their
arrangement in a periodic fashion with distance $a_0$ drastically
reduces the effect of disorder on large scales.
\cite{nattermann1990,korshunov1993,giamarchi1994} \emph{ii})~The
line-nature leads to a tensorial elasticity with the shear modulus
different from the compression and tilt moduli; this introduces a
weak dependence of the wandering exponent upon the ratio of
compression and shear moduli. \cite{bogner2001}
\emph{iii})~Long-range interactions between the vortex lines
stiffen up the lattice as expressed by dispersive
compression and tilt moduli, \cite{brandt1977,larkin1979} thereby
qualitatively reducing the effect of disorder; in the presence of
transverse screening, this effect is limited to scales below the
London penetration length $\lambda$. \emph{iv}) The long-range
interaction between the vortices and the defects implies a
long-range correlation in the disorder landscape, modifying the
physics of thermal depinning.\cite{blatter1994,mueller2001b}

In this paper, we focus on the effect of the long-ranged 
interaction between
the vortices, see \emph{iii}) above. This interaction plays an 
important role
in the short- to intermediate distance regime before being cut 
off by
transverse screening at $R\sim\lambda$ in transverse-, $L\sim 
\lambda^2/a_0$ in
longitudinal direction. Let us put the results for the 
displacement correlator
into context and follow its evolution from small to large scales:

Because of the line-nature of the vortex lattice (cf.\ \emph{ii})
all results are anisotropic; we will discuss the limits in the 
plane
($R$) and along the direction of the vortex lines ($L$) separately
and comment on the general behavior as appropriate.
We assume weak point-like disorder; the effect of the interaction 
changes the
algebraic growth of the displacement correlator $\langle 
u^2(\bv{r}) \rangle$
into a mere logarithmic growth, as can be seen in lowest-order 
perturbation
theory \cite{larkin1979} valid at small distances $R<R_\mrm{c}$ 
and in the weak
pinning limit,
\begin{equation}
   \langle u^2(R) \rangle
   = \xi^2 \frac{\ln \left[R / \alpha a_0\right] }
   {\ln \left[R_\mrm{c}/\alpha a_0\right]}.
   \label{firstlog}
\end{equation}
Here, $\xi$ is the correlation length of the disorder landscape 
properly 
defined 
in Eq.\ (\ref{defxi}) (for point-like disorder the scale $\xi$ is 
of the order 
of 
the vortex core dimension) and $\alpha$ is a number of order 
unity. Later, we 
will 
obtain $\alpha \approx c/\sqrt{\pi}$, where $c = e^{-\gamma}$ and 
$\gamma 
\approx 
0.577$ is the Euler constant. The result $\langle u^2(L)\rangle$
in the longitudinal direction is obtained from the above equation 
via the
substitution $R\to [(c/\sqrt{\pi}) a_0 L]^{1/2}$.  The dependence 
on the 
disorder 
strength has been encoded in the collective bundle pinning radius 
$R_\mrm{c}$ 
defined 
via $\langle u^2(R_\mrm{c}) \rangle = \xi^2$; for
suitably weak disorder we have $a_0 < R_\mrm{c} < \lambda$ and we 
will discuss
other cases later. The long-range nature of the interaction 
between vortices is
relevant for
\begin{equation}
a_0 < R  < \lambda , \qquad a_0<L<\lambda^2/a_0 . 
\label{rbar-disp}
\end{equation}

When $\langle u^2(\bv{r})\rangle$ increases beyond $\xi$ at 
$R_\mrm{c}$ the
appearance of multiple minima leads to a breakdown of the 
perturbative
analysis.\cite{villain1983} For $R_\mrm{c}<R <\lambda$, we use the
renormalization group (RG) to calculate the mean-squared 
displacement and find
the result
\begin{equation}
   \langle u^2(R) \rangle
   \sim \xi^2 \left( \frac{\ln\left[ R/a_0\right]}
   {\ln \left[R_\mrm{c}/a_0\right]} \right)^{2\sigma},
   \label{sublog1}
\end{equation}
exhibiting a sub-logarithmic growth with $\sigma \approx 0.174$ 
(to the
precision assumed here, we drop the numerical $\alpha$ from Eqs.\
(\ref{sublog1}) to (\ref{bglog})); the substitution $R\to 
\sqrt{a_0L}$ again
provides the result along the longitudinal direction. Equation 
(\ref{sublog1})
is the central result of this paper and will be derived below.

On larger length scales, the static properties of the disordered 
flux-line
lattice are well understood. \cite{blatter1994,nattermann2000} At 
a distance
$R_a$ the displacements become of the order of the lattice 
constant $a_0$,
$\langle u^2(R_a)\rangle \sim a_0^2$, and the effect of disorder 
is reduced
since the flux-lines no longer probe independent disorder 
realizations. In the
absence of transverse screening the correlator then has an even 
slower growth
$\langle u^2(R) \rangle \propto \ln \ln [R/R_a]$. 
\cite{chitra1999} In a real
system, however, the vortex-vortex interaction is usually 
screened on scales
$\lambda$ smaller than $R_a$, the lattice gets softer and 
therefore rougher.
For $\lambda<R<R_a$ (random manifold regime) one 
obtains\cite{blatter1994}
\begin{equation}
   \langle u^2(R)\rangle
   \sim \xi^2 \left( \frac{R / \lambda }
   {\ln [R_\mrm{c}/a_0]} \; \right)^{2\zeta}.
   \label{randommanifold}
\end{equation}
The result in longitudinal direction can be obtained via the 
substitution $R\to
a_0L/\lambda$ (the anisotropy is modified as compared to the case 
$R<\lambda$
above). The functional RG calculation\cite{bogner2001} provides a 
value of the
roughness exponent $\zeta$ which depends weakly on the ratio of 
the compression
($c_{11}$) and shear ($c_{66}$) moduli, assuming the value 
$\zeta\approx 0.174$
in the limit $c_{11}\gg c_{66}$ relevant in the vortex lattice. 
Finally,
periodicity becomes important at $R_a$ (Bragg Glass regime) and 
one recovers a
logarithmic growth 
\cite{nattermann1990,korshunov1993,giamarchi1994}
\begin{equation}
   \langle u^2(R) \rangle
   \sim a_0^2\;\ln\left(\frac{R}{R_a}\right),
   \label{bglog}
\end{equation}
where the prefactor of the logarithm does not depend on the 
disorder strength
(the latter is encoded in $R_a$). The physical properties of the 
vortex
lattice, its periodicity and the long-ranged interaction between 
the
constituents, thus qualitatively reduce the effect of disorder.

The paper is organized as follows: In the next section we present
the model and perform the perturbative calculation which provides
the starting point for the renormalization group analysis. Section
III is devoted to the derivation of the flow equation for the
disorder correlator and the calculation of the effective
mean-squared relative displacement. We summarize and conclude in
section IV.

\section{The model}
We consider an isotropic type-II superconductor in the mixed 
state and use a
coordinate system with the magnetic field $B$ applied along the
$\hat\bv{z}$-axis. In the absence of disorder the vortices form a 
triangular
Abrikosov lattice with a lattice constant 
$a_{\scriptscriptstyle\triangle}$,
$a^2_{\scriptscriptstyle\triangle}=2a_0^2/\sqrt{3}$, and $a_0^2 = 
\Phi_0/B$
($\Phi_0=h c/2e$ is the magnetic flux quantum). The disordered 
vortex system is
described by a free energy involving both elastic and pinning 
parts. Within the
continuum elastic theory, the elastic part takes the form 
\cite{blatter1994}
\begin{eqnarray}
    \mathcal{F}_\mrm{el} [\bv{u}] = &&
    \!\!\!\!\!\!\frac{1}{2} \int
    \frac{d^3k}{(2\pi)^3}  \left[
    c_{11}(\bv{k}) |\bv{K \cdot u(k)}|^2 \right.
    \label{elastic-energy}\\
    &&\left.+ c_{66}|\bv{K_\bot \cdot u(k)}|^2 \nonumber
    + c_{44}(\bv{k}) k_z^2 |\bv{u(k)}|^2 \right],
\end{eqnarray}
where $\bv{k}=(\bv{K},k_z)=(k_x,k_y,k_z)$, $\bv{K}_\bot = 
(k_y,-k_x)$, and with
the integration defined over the lattice Brillouin zone. The 
compression and
tilt moduli are strongly dispersive, i.e., they depend on the 
wave vector
\bv{k} of the distortion,
\begin{equation}
   c_{11}(\bv{k}) \approx c_{44}(\bv{k})
   \approx \frac{\hat{c}_{44}}{1+\lambda^2k^2},
   \quad \hat{c}_{44} = \frac{B^2}{4\pi},
\end{equation}
where $k=|\bv{k}|$. The dispersion is due to the long-range 
current-mediated
interaction between the vortices and is relevant in the 
intermediate distance
regime,
\begin{equation}
a_0 < R  < \lambda , \qquad a_0<L<\lambda^2/a_0
\end{equation}
($\bv{r}=(\bv{R},L)$, $R=|\bv{R}|$), before it is cut on the 
scale of the
penetration depth $\lambda$; we will restrict ourselves to this 
intermediate
regime in the following. As usual, we ignore the weak dispersion 
in the shear
modulus $c_{66}$,
\begin{equation}
   c_{66} \approx \frac{\Phi_0B}{(8\pi\lambda )^2}.
\end{equation}
The static Green function corresponding to the elastic energy
(\ref{elastic-energy}) reads ($\mu , \nu=x,y$)
\begin{eqnarray}
 G^{\mu\nu}(\bv{k})&=&\frac{\mathcal{P}_{\mrm{L}}^{\mu\nu}(\bv{K
})}
   {c_{11}(\bv{k}) K^2 + c_{44}(\bv{k})k_z^2}
   \label{greenfunction} \\
   &&\qquad+\frac{\mathcal{P}_{\mrm{T}}^{\mu\nu}(\bv{K})}
   {c_{66} K^2 + c_{44}(\bv{k})k_z^2}, \nonumber
\end{eqnarray}
with the longitudinal and transverse projection operators
$\mathcal{P}_{\mrm{L}}^{\mu\nu}(\bv{K})=K_\mu K_\nu /K^2$ and
$\mathcal{P}_{\mrm{T}}^{\mu\nu}(\bv{K})= \delta^{\mu\nu}
-\mathcal{P}_{\mrm{L}}^{\mu\nu}(\bv{K})$, respectively ($K\equiv
|\bv{K}|$). In the regime we are interested in, the compression
modes involve large energies ($c_{11}\gg c_{66}$) and the
longitudinal part of the Green function can be neglected; we
denote the transverse part by $G_\mrm{T}^{\mu\nu}$.

Assuming weak point-like defects, we describe the disorder term in
the free energy via the pinning energy density $E_\mrm{pin}
(\bv{r},\bv{u(r)})$,
\begin{equation}
   \mathcal{F}_\mrm{pin} [\bv{u}]
   \equiv \int d^3r \; E_\mrm{pin} (\bv{r},\bv{u(r)}).
\end{equation}
The pinning energy density $E_\mrm{pin}$ derives from the
convolution of the pinning potential with the vortex form function
\cite{blatter1994} and is characterized by the disorder
correlator,
\begin{equation}
   \langle E_\mrm{pin}(\bv{r,u}) E_\mrm{pin}(\bv{r',u'}) \rangle
   = \delta^3(\bv{r-r'}) K_0(\bv{u-u'}),
\end{equation}
where $K_0$ is proportional to the disorder strength and is a 
short-range
function decaying over the characteristic size $\xi$ of the 
vortex core. For
the disordered vortex lattice the correlator has an algebraic
decay,\cite{blatter1994} $K_0 \propto (1/u^2) \ln(u/\xi)$ for 
$u<\lambda$.
We consider weak disorder where the vortices are pinned in 
bundles of size
$R_\mrm{c} > a_0$; this contrasts with single vortex pinning 
realized for
larger disorder strength to be discussed later.
The above description remains valid as long as
the disorder-induced displacements remain small, $\langle 
u^2\rangle<
a^2_0$, such that the vortices probe a different random 
environment upon
deformation. As the displacement becomes of the order of $a_0$ on 
large scales,
one has to take the periodicity of the lattice into account.

The equilibrium configuration $\bv{u(r)}$ of the system is
obtained by minimizing the free energy functional $\mathcal{F}
[\bv{u}] = \mathcal{F}_\mrm{el} + \mathcal{F}_\mrm{pin}$ and is
implicitly given by
\begin{equation}
   u^\mu (\bv{r})
   = - \int d^3r' \; G^{\mu\nu}(\bv{r-r'})\;
   \frac{\partial}{\partial u^\nu} E_\mrm{pin}(\bv{r',u(r')}),
   \label{implicit-u}
\end{equation}
where $G^{\mu\nu}(\bv{r})$ is the Fourier transform of the Green
function (\ref{greenfunction}). This `equation of state' allows
for calculating the mean-squared displacement correlator $\langle
u^2(\bv{r})\rangle $.

\subsection{Perturbation theory}
For small displacements $\langle u^2(\bv{r})\rangle <\xi^2$, the 
\bv{u}
dependence in the force ($-\partial_{u^\nu}E_\mrm{pin}$) can be 
neglected in
(\ref{implicit-u}), and the mean-squared relative displacement 
can be written
in the form
\begin{eqnarray}
   \langle u^2(\bv{r}) \rangle
   &\approx & -2 K_0^{\mu\nu}(0) \int \frac{d^3k}{(2\pi)^3}
   (1-\cos \bv{k r}) \nonumber \\
   && \times G^{\rho\mu}(\bv{k})G^{\rho\nu}(\bv{-k}),
   \label{usquared}
\end{eqnarray}
with the integration restricted to the lattice Brillouin zone in 
the plane.

We denote derivatives of the correlator with respect to $u^\nu$ by
a superscript,
\[
   K_0^\nu \equiv \frac{ \partial}{\partial u^\nu} K_0,
\]
and sum over indices that appear twice. Equation (\ref{usquared}) 
simplifies
since $K_0$ is isotropic: $K_0^{xx}(0)=K_0^{yy}(0)=K_0''(0)$ and 
$K_0^{xy}(0)=0$
(the primes denote derivatives of $K_0(u)$ with respect to 
$u\equiv |\bv{u}|$).
Inserting the transverse part of the Green function 
(\ref{greenfunction}), we
obtain
\begin{equation}
   \langle u^2(\bv{r}) \rangle
\approx   -2 K_0''(0) \int \frac{d^3k}{(2\pi)^3}
   \frac{1- \cos \bv{k r}}{[c_{66} K^2 + c_{44}(k)k_z^2]^2}.
\end{equation}

In the intermediate regime, $a_0<R<\lambda$, 
$a_0<L<\lambda^2/a_0$, the
relevant values of $k_z$ are much smaller than $K$. Hence, we can 
approximate
$c_{44}(\bv{k})=\hat c_{44}/(1+\lambda^2k^2)\approx \hat 
c_{44}/\lambda^2K^2$
and obtain
\begin{equation}
   \langle u^2(\bv{r}) \rangle
   \approx  - 2 K_0''(0) \int\frac{d^3k}{(2\pi)^3}
   \frac{[1-\cos(\bv{kr})]\lambda^4 K^4 }{[c_{66}\lambda^2K^4
   + \hat{c}_{44}k_z^2]^2} . \label{usq}
\end{equation}

The divergence of the Green function as $\bv{k}\to 0$ is cut off 
by the
numerator on the scale $k\sim 1/r$. For large wave vectors, the 
integral is
limited by $k\sim 1/a_0$ (bundle pinning regime). In a first 
analysis, we
therefore replace (\ref{usq}) by
\begin{equation}
   \langle u^2(\bv{r}) \rangle
   \approx -\frac{1}{2} K_0''(0) \; I \frac{2}{\pi^2}
   \int d^2Kdq \frac{K^4}{[K^4+q^2]^2},
   \label{simpleintegral}
\end{equation}
where $q\equiv k_z\sqrt{\hat c_{44}/c_{66}}/ \lambda
\approx k_z 4 \sqrt{\pi} /a_0$ and
\begin{equation}
   I\equiv \frac{\lambda }{4\pi c_{66}\sqrt{c_{66}\hat c_{44}}}.
\end{equation}
The integral in (\ref{simpleintegral}) is essentially 
four-dimensional,
\[
   \frac{2}{\pi^2}\frac{K^4d^2Kdq}{[K^4 +q^2]^2}
   \hat = \frac{1}{2\pi^2} \frac{d^3(K^2)dq}{ [(K^2)^2+q^2]^2}
    = \frac{1}{2\pi^2} \frac{d^4y}{y^4} \hat = \frac{dy}{y}
\]
(the symbol $\hat =$ reminds that we are using $d^dx = x^{d-1} 
S_d \, dx$, with
$S_d$ the surface of the $d$-dimensional sphere), and has a 
logarithmic
divergence at small $y\equiv (K^4+q^2)^{1/2}$. The integral
(\ref{simpleintegral}) is dominated by values $q \sim K^2$; for 
$R^2\gg a_0 L$
the small scale cut-off is given by $y\sim 1/R^2$, while a 
cut-off $y \sim
1/a_0L$ applies in the opposite limit $L \gg R^2/a_0$. We 
therefore
obtain\cite{blatter1994}
\begin{equation}
   \langle u^2(\bv{r}) \rangle
   \approx   - K_0''(0)\; I \ln\left(\frac{\rho(\bv{r})}
   {a_0 }\right) ,\label{roughusq}
\end{equation}
where the function $\rho(\bv{r})$ assumes the limits
\begin{eqnarray}
\rho(R^2\gg a_0L)& = & R/\alpha \; , \nonumber \\ \rho(R^2 \ll 
a_0L)& =&
\left(a_0L\right)^{1/2}/\alpha' \; , \label{rho}
\end{eqnarray}
with $\alpha$ and $\alpha'$ numbers of order unity; the proper 
interpolation
for arbitrary values of $R$ and $L$ is beyond the accuracy of the 
present 
analysis.
Note that as a consequence
of the dispersion in the tilt modulus, the lengths along the 
longitudinal ($L$)
and transverse ($R$) directions scale differently, $L\sim R
\sqrt{c_{44}/c_{66}} \sim R^2/a_0$ for $R<\lambda$, 
$L<\lambda^2/a_0$. The
logarithmic growth of (\ref{roughusq}) is a consequence of the 
dispersion in
the tilt modulus (and hence of the long-range interaction), 
effectively lifting
the problem to four dimensions where the disorder is only 
marginally
relevant;\cite{blatter1994,nattermann2000} the prefactor of the 
logarithm is
proportional to the disorder strength.

The simplifications made in (\ref{simpleintegral}) do not allow 
for a
determination of the numericals $\alpha$, $\alpha'$, and the shape
$\rho(\bv{r})$; in the following we perform a more accurate 
calculation
of the original expression (\ref{usquared}) in order to obtain a 
precise
result for the argument under the logarithm in (\ref{roughusq}).
We approximate the hexagonal lattice
Brillouin zone by the circular version 
$K<K_{\rm\scriptscriptstyle BZ} =
\sqrt{4\pi}/a_0$; using cylindrical coordinates 
$\bv{k}=(K,\phi,k_z)$, the
angular integration in (\ref{usq}) can be carried out directly 
and the
integration over the $k_z$-axis can be performed using residues, 
after which
\begin{equation}
   \frac{\langle u^2(\bv{r}) \rangle}{I\,K_0''(0)}
   \approx  -
    \!\int \limits_0^{K_{\rm\scriptscriptstyle BZ}}\!\! dK
     \frac{1-J_0(KR) e^{-a_0 L K^2/4\sqrt{\pi}} }{K} .
     \label{simplifiedintegral}
\end{equation}
For $R$ or $L$ are much larger than $a_0$, the integral in
(\ref{simplifiedintegral}) can be approximated by
\[
 A(R,L) \equiv
\int_0^\infty \!\!\!\! dK
     \frac{1-J_0(KR) e^{-a_0 L K^2/4\sqrt{\pi}}}{K}\;
     e^{-c K^2 /K^2_{\rm\scriptscriptstyle BZ}},
\]
where the constant $c$ is fixed by demanding
\[ \int_\delta^{K_{\rm\scriptscriptstyle BZ}} \frac{dK}{K}
= \int_\delta^\infty \frac{dK}{K}\; e^{-cK^2 
/{K^2_{\rm\scriptscriptstyle BZ}}}
\]
for $\delta\to 0$; one obtains $c=e^{-\gamma}\approx 0.56$, where 
$\gamma
\approx 0.577$ is the Euler constant.

The integral $A(R,L)$ exhibits a logarithmic behavior for small 
$K$ which then
is cut off either by $R$ or by $\sqrt{La_0}$. The approximation 
of a circular
Brillouin zone (see above) is correct up to a factor of order 
unity under
the logarithm.

We directly evaluate the integral $A(R,L)$ and obtain
\begin{eqnarray}
A(R,L) &=& \ln \biggl(\frac{R}{(c/\sqrt{\pi})a_0} \; \biggr) 
\label{ARL}
%\\&&\qquad
+ \frac{1}{2} \; \Gamma\biggl( 0,\frac{\pi R^2}{c a_0^2 + 
\sqrt{\pi}a_0 L}
\biggr) \nonumber
\end{eqnarray}
($\Gamma(0,x)=\int_x^\infty dt\, e^{-t}/t$ is the incomplete 
gamma function),
displaying a complicated dependence on the cut-off parameters $R$ 
and $L$. In
the limit of large $R$, $R^2 \gg 2ca_0^2/\pi + 2a_0 
L/\sqrt{\pi}$, the above
result simplifies to
\begin{equation}
A(R,L) \approx \ln \biggl(\frac{R}{(c/\sqrt{\pi}) a_0}\; \biggr), 
\label{R>L}
\end{equation}
while in the opposite limit of large $L$, $R^2 \ll 2ca_0^2/\pi
+ 2a_0 L/\sqrt{\pi}$ (and $L\gg c a_0/\sqrt{\pi}$) one obtains
\begin{equation}
A(R,L) \approx \ln \biggl(\frac{L}{(c/\sqrt{\pi}) a_0} \; 
\biggr)^{1/2}.
\label{R<L}
\end{equation}
The present, more accurate, analysis then allows for the 
definition
of the cut-off function $\rho(R,L)$ for arbitrary values of $R$ 
and $L$
within the perturbative regime,
\begin{equation}
\rho(R,L) = a_0 \, e^{A(R,L)} .\label{rbar}
\end{equation}
Comparing the above results with (\ref{roughusq}) and 
(\ref{rho}), we find the
numericals $\alpha \approx c/ \sqrt{\pi} \approx 0.32$ and
$\alpha'=\alpha^{1/2}\approx 0.56$. Defining the collective 
pinning radius 
$R_\mrm{c}$ 
via
\begin{equation}
\langle u^2(R_\mrm{c}) \rangle = \xi^2 ,
   \label{urc}
\end{equation}
we obtain (\ref{firstlog}) with the given value of the numerical 
$\alpha$.

The above perturbative results are valid within the collective 
pinning regime
$R<R_\mrm{c}$, which is equivalent to $\rho(R,L) < 
R_\mrm{c}/\alpha \approx 
3.16\,R_\mrm{c}$.
Beyond this regime, the perturbative result (\ref{roughusq})
breaks down because of the appearance of multiple minima as the 
displacement
$\langle u^2(\bv{r})\rangle^{1/2}$ increases beyond the 
characteristic scale
$\xi$ of the disorder potential.\cite{villain1983} Along the 
longitudinal
direction, this breakdown occurs at the larger scale 
$L_\mrm{c}^\mrm{b} \approx
3.16\, R_\mrm{c}^2/a_0>R_\mrm{c}$. Within the present 
perturbative approach the
length scale $\xi$ appearing in (\ref{urc}) has been introduced 
{\it ad hoc}
--- this will be different in the functional RG treatment 
extending the
discussion to larger length scales beyond $R_\mrm{c}$.

\section{functional RG equation}

The logarithmic behavior of the displacement correlator 
(\ref{roughusq})
provides the motivation for applying the renormalization group
\cite{fisher1986} (RG); we briefly summarize the main steps in 
the derivation
of the functional RG equation for the disorder correlator 
$K_0(u)$ using the
real-space renormalization procedure introduced in Ref.\
\onlinecite{bucheli1998}.  We aim at finding the effective 
disorder correlator
on large scales ($R_\mrm{c} <R <\lambda$, 
$L_\mrm{c}^\mrm{b}<L<\lambda^2/a_0$).
The elastic coefficients are not renormalized as the pinning part 
of the system
is invariant under the `tilt' $\bv{u(k)} \to 
\bv{u(k)}+\bv{v}(\bv{k})$ for any
vector-function $\bv{v}$. \cite{schulz1988,balents1993}

Assuming renormalization up to a distance $(R_1,L_1)$, we define 
the
displacement $\bv{u}_1$, the energy 
$E_\mrm{pin}^{\scriptscriptstyle (1)}$, and
the correlator $K_1$ at the scale $(R_1,L_1)$. We perform the RG 
step to
$R_2>R_1$, $L_2>L_1$, and split $\bv{u}_1=\bv{u}_2+\bv{w}$ (using 
the self
consistency equation (\ref{implicit-u})) into a far-
\[
   u_2^\mu (\bv{r}) =
   -\int_{\Omega^>} d^3r' \;
   G^{\mu\nu}(\bv{r-r'}) \partial_{u^\nu}
   E^{(1)}_\mrm{pin} (\bv{r',u}_1\bv{(r')})
\]
with $\Omega^> =\{|\bv{R-R}'|>R_2;\; L-L'>L_2\}$, and a near-field
contribution,
\[
   w^\mu (\bv{r})
   = -\int_\Omega
   d^3r' \; G^{\mu\nu}(\bv{r-r'})\partial_{u^\nu}
   E^{(1)}_\mrm{pin} (\bv{r',u}_1\bv{(r')}),
\]
where $\Omega=\{R_1<|\bv{R-R}'|<R_2;\; L_1<L-L'<L_2 \}$.
We express the effective free energy 
$E_\mrm{pin}^\mrm{\scriptscriptstyle (2)}
(\bv{u}_2(\bv{r),r})$ for the far-field contribution as a Taylor 
expansion in
the near-field $\bv{w(r)}$ and, assuming Gaussian disorder, 
determine the
effective pinning energy correlator
\[
   K_2(\bv{u}_2-\bv{u}_2')\delta^3(\bv{r-r'})
   \equiv \langle E_\mrm{pin}^\mrm{(2)}(\bv{r,u}_2)\;
   E_\mrm{pin}^\mrm{(2)}(\bv{r'},\bv{u}_2')
\rangle
\]
on the scale $(R_2,L_2)$. Evaluating the terms to second order in 
$K_1$ (first
loop), one obtains
\begin{equation}
   K_2(\bv{u}_2)
   = K_1(\bv{u}_2) + I^{\mu\nu\rho\kappa}
   \left[ \frac{1}{2} K_1^{\mu\rho}K_1^{\nu\kappa}
   - K_1^{\mu\kappa}(0)K_1^{\nu\rho} \right],
   \label{Keff}
\end{equation}
where
\begin{equation}
   I^{\mu\nu\rho\kappa}
   \equiv \int_\Omega d^3r' \; G^{\mu\nu}(\bv{r-r}')
   G^{\rho\kappa}(\bv{r-r}').
   \label{integral}
\end{equation}
The Green function is most conveniently expressed in Fourier 
space,
\begin{equation}
   I^{\mu\nu\rho\kappa}
   = \int_{\Omega'} \frac{d^3k}{(2\pi)^3} \; G^{\mu\nu}(\bv{k})
   G^{\rho\kappa}(\bv{k}), \label{I_k}
\end{equation}
where $\Omega'$ is the appropriate $k$-space integration domain 
corresponding 
to 
$\Omega$
(see below). Since $c_{11} \gg c_{66}$ we can neglect the 
longitudinal part of
the Green function (\ref{greenfunction}) and inserting the 
transverse part
$G_\mrm{T}$ into (\ref{integral}) the tensorial structure of the 
integral can
directly be calculated using cylindrical coordinates 
$(K,\phi,k_z)$,
\begin{equation}
   \int_0^{2\pi} d\phi \; \mathcal{P}_{\mrm{T}}^{\mu\nu}(\bv{K})
   \mathcal{P}_{\mrm{T}}^{\rho\kappa}(\bv{K})
   = \frac{2\pi}{8} \Delta^{\mu\nu\rho\kappa},
   \label{tensorialstructure}
\end{equation}
where
\begin{equation}
   \Delta^{\mu\nu\rho\kappa}
   \equiv \delta^{\mu\nu}\delta^{\rho\kappa}
   + \delta^{\mu\rho}\delta^{\nu\kappa}
   + \delta^{\mu\kappa}\delta^{\nu\rho}
\end{equation}
(for the isotropic situation with $G^{\mu\nu}\propto 
\delta^{\mu\nu}$ one
obtains $I^{\mu\nu\rho\kappa} \propto 
\delta^{\mu\nu}\delta^{\rho\kappa}$).

Equation (\ref{I_k}) resembles the equation for the displacement
correlator (\ref{usquared}); following Ref.\ 
\onlinecite{wagner1999} we 
approximate
the term $1+\lambda^2k^2$ in $c_{44}(\bv{k})$ by $\lambda^2K^2$ 
and use the
fact, that the resulting integral is effectively 
four-dimensional, cf.\
(\ref{simpleintegral}). Performing the RG-step only in transverse 
direction
($L_2=L_1$), the $k_z$-integration extends over the whole axis, 
while
the $K$-integration is cut-off at $K_2\sim 1/R_2$ and $K_1\sim 
1/R_1$; in the
opposite case, the integral is cut by $k_{z2}\sim 1/L_2$ and 
$k_{z1}\sim 1/L_1$.
The final result is conveniently expressed\cite{wagner1999} 
through the
function $\rho(\bv{r})$ as defined in (\ref{rho})
\begin{equation}
   I^{\mu\nu\rho\kappa}
   = \frac{I}{16} \Delta^{\mu\nu\rho\kappa}
   \ln \left(\frac{\rho(R_2,L_2)}{\rho(R_1,L_1)}\right);
   \label{I-alphabeta}
\end{equation}
note that, given the accuracy of the present analysis, we do not 
make
use of the result (\ref{rbar}) which has been derived within the
perturbative regime only. We define the flow parameter
\begin{equation}
l\equiv\ln [ \rho(R,L) / a_0],
\end{equation}
where the starting point of the flow is chosen by matching our RG 
analysis with
the result from perturbation theory, cf.\ (\ref{roughusq}).

Inserting (\ref{I-alphabeta}) into (\ref{Keff}) yields the RG 
equation for the
effective pinning energy correlator,
\begin{equation}
   \partial_l K_l(\bv{u})
   = \frac{I}{16} \Delta^{\mu\nu\rho\kappa}
   \left(\frac{1}{2} K_l^{\mu\rho}K_l^{\nu\kappa}
   - K_l^{\mu\kappa}(0)K_l^{\nu\rho}\right). \label{rg-for-k}
\end{equation}
Summing over pairs of indices, this simplifies to
\begin{eqnarray}
   \partial_l K_l
   = \frac{I}{16}\bigg( \!\!\!\!\!\!\!&& K_l^{\mu\nu} K_l^{\mu\nu}
   -2 K_l^{\mu\nu} K_l^{\mu\nu}(0)+ \nonumber \\
   && \frac{1}{2} K_l^{\mu\mu} K_l^{\nu\nu}
   -K_l^{\mu\mu} K_l^{\nu\nu}(0)\bigg).
   \label{generalrgeq}
\end{eqnarray}
The first line of this RG equation contains the terms that are 
obtained for an
isotropic situation\cite{halpin-healy1989,balents1993} with 
$G^{\mu\nu}\propto
\delta^{\mu\nu}$; the terms in the second line arise due to the 
transverse
structure of the Green function entering 
(\ref{tensorialstructure}). Up to
scaling terms, an equation equal to (\ref{generalrgeq}) is 
obtained in the
non-dispersive [3+2]-case for $c_{11}\gg c_{66}$, while a 
different tensorial
structure is found assuming arbitrary values of $c_{11}$ and 
$c_{66}$, cf.\
Ref.\ \onlinecite{bogner2001}.

In the regime we are interested in, the correlator is isotropic,
$K_l(\bv{u})=K_l(u)$, and the derivatives can be rewritten as
\begin{eqnarray}
   K_l^{\mu\nu}(\bv{u})
   &=& \left(K_l''(u) - \frac{K_l'(u)}{u} \right)
   \frac{u^\mu u^\nu}{u^2}
   + \frac{K_l'(u)}{u} \delta^{\mu\nu}, \nonumber \\
   K_l^{\mu\nu} (0) &=& K_l''(0) \delta^{\mu\nu};
\end{eqnarray}
the RG equation (\ref{generalrgeq}) then takes the form
\begin{eqnarray}
   \partial_l K_l(u)
   = \frac{I}{16} \bigg[\!\!\!\!\!\!\!&&
   \frac{3}{2}\left[K_l''\right]^2 + K_l''\frac{K_l'}{u}
   + \frac{3}{2}\left(\frac{K_l'}{u}\right)^2 \nonumber \\
   &&  -4K_l''(0) \left(K_l'' + \frac{K_l'}{u} \right) \bigg],
   \label{sphericalrgequation}
\end{eqnarray}
where the primes denote derivatives with respect to $u=|\bv{u}|$. 
Taking four
derivatives with respect to $u$ and assuming an analytic 
correlator, one
obtains
\[ \partial_l K_l^{(4)} (0)= I [K_l^{(4)}(0)]^2. \]
At the scale $l_\mrm{c}$, the fourth derivative of the correlator 
becomes
infinite at the origin, $K^{\scriptscriptstyle 
(4)}_{l>l_\mrm{c}}(0) =\infty$,
and the second derivative develops a cusp singularity. This 
non-analyticity
provides a definition for the collective pinning radius 
$R_\mrm{c}$ which
depends on the fourth derivative $K_0^{\scriptscriptstyle 
(4)}(0)$ of the
initial correlator,
\begin{equation}
   l_\mrm{c} = \frac{1}{IK_0^{(4)}(0)} \equiv
   \ln \left[ R_\mrm{c}/\alpha a_0\right].
   \label{lc}
\end{equation}

Next, we determine the static structure of the vortex lattice on 
large scales.
As for the disorder correlator $K_l$, we use perturbation theory 
to find the
renormalized displacement field $\langle u^2(\bv{r})\rangle$ on 
scale $\bv{r}$.
We split the integral (\ref{simpleintegral}) into shells and 
replace $K_0''(0)$
in each shell by the scale-dependent disorder correlator 
$K_l''(0)$.
Integrating over all shells we obtain
\begin{equation}
\langle u^2(\bv{r}) \rangle
   \approx   - \; I\int_{\ln a_0}^{\ln \rho(\bv{r})}\!\!dl'\,
   K_{l'}''(0)\; . \label{RGeqforu}
\end{equation}
As in (\ref{simpleintegral}), the integral is cut off by the 
function $\rho
(\bv{r})$, which is defined in (\ref{rho}) via its limits for 
$R^2$ much larger
or much smaller than $a_0L$.

\section{Analysis of the RG flow}
Analyzing (\ref{sphericalrgequation}), the disorder correlator 
$K_l$ flows
towards zero on large scales, indicating that the disorder is 
only marginally
relevant. Here, we are interested in learning how $K_l''(0)$ 
flows to zero with
increasing $l$ in order to integrate the equation 
(\ref{RGeqforu}) for the
displacement correlator. We will see that a proper rescaling of 
the
displacement field $\bv{u(r)}$ and of the correlator $K_l$ maps 
the flow
equation (\ref{sphericalrgequation}) describing the marginal 
situation to the
flow equation in one dimension less \cite{emig1998,chitra1999}; 
this will allow
us to make use of results derived from an $\epsilon = 
4-d$-expansion at the
value $\epsilon = 1$.

Following (\ref{RGeqforu}) we need to integrate the flow equation
for $K_l''(0)$. Differentiating (\ref{generalrgeq}) or
(\ref{sphericalrgequation}) twice with respect to $u$ and
evaluating the result at $u=0$ one obtains
\begin{equation}
   \partial_l K_l''(0) = \frac{3I}{16} K_l'''(0^+)^2 
.\label{k3strich}
\end{equation}
In the Larkin regime, $R <R_\mrm{c}$, $L<L_\mrm{c}^\mrm{b}$, the 
disorder
correlator is an analytic function ($K_l'''(0^+)=0$) and hence
$K_{l<l_\mrm{c}}''(0)=K_0''(0)$ remains constant; integrating 
(\ref{RGeqforu})
we confirm the perturbative result (\ref{roughusq}),
\begin{equation}
   \langle u^2(\bv{r}) \rangle
   = I \int_0^{l<l_\mrm{c}} dl' [-K_{l'}''(0)]
   = -K_0''(0)\; I \; l,
   \label{usRG}
\end{equation}
with $l=\ln[\rho(R,L) /\alpha a_0]$. This result is in agreement 
with the
statement in Ref.\ [\onlinecite{efetov1979}] that higher order 
terms in
perturbation theory do not contribute to $\langle u^2(\bv{r}) 
\rangle$.

Combining (\ref{lc}) and (\ref{usRG}) at $\rho=\rho(R_\mrm{c})$, 
we find the
proper definition of the relevant length scale $\xi$ of the 
disorder landscape
in terms of derivatives of $K_0$,
\begin{equation}
    \langle u^2(R_\mrm{c}) \rangle
    = \frac{-K_0''(0)}{K_0^{(4)}(0)}
    \equiv \xi^2;
    \label{defxi}
\end{equation}
this precise definition of $\xi$ naturally appears within the 
renormalization
group framework.

Going beyond $l_c$, it is more convenient to investigate the 
asymptotic flow of
$K_l(u)$, rather than solving Eq.\ (\ref{k3strich}). Since the 
magnitude
$K_l(0)$ decreases and the function $K_l(u)$ broadens, we perform 
the rescaling
\begin{equation}
   \frac{I}{16}K_l(u)
   = \frac{1}{l^\tau}f_l\left(\frac{u}{l^\sigma}\right)
   \label{ansatz}
\end{equation}
in order to extract the characteristic geometry of the correlator 
(its height
and width). The simplified flow equation for $f_l$ then will 
converge to a
regular and finite function $f^*$ for large $l$ such that we 
arrive at a proper
asymptotic scaling form for $K_l \sim l^{-\tau} f^*(u/l^\sigma)$. 
Inserting
(\ref{ansatz}) into (\ref{sphericalrgequation}), the resulting 
flow equation
involves terms proportional to $l^{-\tau-1}$ and 
$l^{-2\tau-4\sigma}$; we
require these terms to scale the same way in order to achieve a 
flow $f_l \to
f^*$ and hence we fix $\tau=1-4\sigma$. The flow equation for 
$f_l$ then reads
\begin{eqnarray}
   \frac{\partial}{\partial \ln l} f_l
   &=&  (1-4\sigma ) f_l + \sigma \bar u f_l'
   + \frac{3}{2} [f_l'']^2 \label{eqforf}  \\
   && + f_l''\frac{f_l'}{\bar u}
   + \frac{3}{2}\left( \frac{f_l'}{\bar u}\right)^2
   -4f_l''(0) \left( f_l'' + \frac{f_l'}{\bar u} \right),
   \nonumber
\end{eqnarray}
where the first two terms on the rhs arise due to the explicit 
$l$-scaling in
(\ref{ansatz}) and the primes denote derivatives with respect to 
$\bar u \equiv
u/l^\sigma$. As we will see, this equation is the same (up to 
$\ln l \to l$) as
the (first loop) flow equation for the rescaled disorder 
correlator $R_l$ in
the non-dispersive [3+2]-case in an  $\epsilon=4-d$-expansion, 
cf.\ Eq.\
(\ref{eqforR}) below.

In the dispersive [3+2]-case, the system is at the upper
critical dimension and the relevant integral for 
$I^{\mu\nu\rho\kappa}$, cf.\
(\ref{integral}),
exhibits a logarithmic behavior, cf.\ (\ref{I-alphabeta}). This 
leads to the
flow equation (\ref{rg-for-k}) in the form $\partial_l K_l ={\cal
O}_{\rm\scriptscriptstyle 1-loop}  (K_l^2)$. In the 
non-dispersive case the
upper critical dimension is four.\cite{blatter1994} In 
$d=4-\epsilon$ the
integral (\ref{integral},\ref{I-alphabeta}) has an algebraic 
behavior
\begin{equation}
   I^{\mu\nu\rho\kappa}
   = \frac{ \tilde I}{16} \Delta^{\mu\nu\rho\kappa}
   \frac{1}{\epsilon} \left(\rho(\bv{r}_2)^\epsilon
   - \rho(\bv{r}_1)^\epsilon \right)
\end{equation}
(with $\tilde I=2I/\pi\lambda$ for $\epsilon =1$), leading to the 
flow equation
$\partial_{\rho^\epsilon} K_{\tilde l} = {\cal 
O}_{\rm\scriptscriptstyle
1-loop} (K_{\tilde l}^2)$, where the flow parameter is
\begin{equation}
   \tilde l = \ln \left( \frac{\rho(R,L)}{\lambda} \right)
\end{equation}
(here, $c_{44} \approx \hat{c}_{44}$ and therefore $\rho(R,L)$ is 
different
from (\ref{rho}): $\rho (R,0)\sim R$, $\rho (0,L) \sim 
a_0L/\lambda$).
Next, we rescale the correlator $\tilde K_{\tilde l} = 
\exp(-\epsilon \tilde
l)\, K_{\tilde l}$ and obtain the flow equation in the form 
$\partial_{\tilde
l} \tilde K_{\tilde l}(u) = \epsilon \tilde K_{\tilde l} + {\cal 
O}_{{\rm
\scriptscriptstyle 1-loop}} (\tilde K_{\tilde l}^2)$.
Finally, we perform the rescaling
\begin{equation}
   \frac{\tilde I}{16}\tilde K_{\tilde l}(u)
   = e^{-\alpha \tilde l} \, R_{\tilde l}
   \left(u e^{-\zeta \tilde l}\right),
   \label{ansatz2}
\end{equation}
corresponding to (\ref{ansatz}) above if we replace $l \to 
\exp(\tilde l)$; the
latter accounts for the change from a logarithmic to an algebraic 
behavior when
going away from the marginal dimension. Carrying out the above 
sequence of
steps, we arrive at the standard flow equation \cite{bogner2001} 
(we define
$\tilde u = u \exp(-\zeta \tilde l)$ and assume $c_{11} \gg 
c_{66}$; the primes
denote derivatives with respect to $\tilde u$)
\begin{eqnarray}
   \frac{\partial}{\partial \tilde l} R_{\tilde l}
   &=&  (\epsilon-4\zeta ) R_{\tilde l}
   + \zeta \tilde u R_{\tilde l}'
   + \frac{3}{2} [R_{\tilde l}'']^2 \label{eqforR} \\
   && + R_{\tilde l}''\frac{R_{\tilde l}'}{\tilde u}
   + \frac{3}{2}\left( \frac{R_{\tilde l}'}{\tilde u}\right)^2
   -4R_{\tilde l}''(0) \left( R_{\tilde l}''
   + \frac{R_{\tilde l}'}{\tilde u} \right),
   \nonumber
\end{eqnarray}
where we have used $\alpha = -4\zeta$. For $\epsilon = 1$, this 
equation is
identical to (\ref{eqforf}) if we replace $\partial \ln l 
\rightarrow \partial
\tilde l$ and $\sigma \rightarrow \zeta$. Comparing to standard 
derivations of
flow equations of the type (\ref{eqforR}), cf.\ Refs.\ 
\onlinecite{balents1993}
and \onlinecite{bogner2001}, here, we have delayed all the 
rescaling of the
field ($u$) and of the coupling function ($K_{\tilde l}$) to the 
very end of
the calculation. The rescaling (\ref{ansatz2}) then again is 
motivated by the
desire for a flow equation for $R_{\tilde l}({\tilde u})$ 
admitting a fix-point
function $R^*({\tilde u})$ corresponding to 
$\zeta^{\scriptscriptstyle 
1L}_{3,2} 
\approx 0.174$ as found in Ref.\ \onlinecite{bogner2001} for 
$c_{11} \gg 
c_{66}$ (we 
denote by $\zeta_{d,N}^{\scriptscriptstyle 1L}$ the one-loop 
value of the 
wandering 
exponent for the $[d+N]$-model).

We briefly remind the arguments of Ref.\ \onlinecite{balents1993}
providing the proper fix-point function $R^*({\tilde u})$ and the
wandering exponent $\zeta$. Depending on the starting conditions,
the numerical integration of the fix-point equation for $R^*$
provides three classes of decaying solutions: for $\zeta <
\zeta^{\scriptscriptstyle 1L}_{3,2}$ the function $R^*(\tilde u)$
crosses the $\tilde u$-axis at least once. Choosing $\zeta >
\zeta^{ \scriptscriptstyle 1L}_{3,2}$, the solution is always
positive and decays in a power law fashion for large $\tilde u$.
Finally, for $\zeta = \zeta^{ \scriptscriptstyle 1L}_{3,2}$,
$R^*({\tilde u})$ is positive as well and decays exponentially.
The first class of solutions is forbidden by the flow as a
positive initial correlator remains positive under the RG flow,
hence $\zeta \geq \zeta^{\scriptscriptstyle 1L}_{3,2}$. On the
other hand, any algebraic tail $R_{\tilde l}(\tilde u)\sim
A_{\tilde l}/{\tilde u}^\gamma$ vanishes under the flow,
\[
\frac{\partial}{\partial \tilde l} A_{\tilde l} =
(1-4\zeta-\gamma\zeta)A_{\tilde l} \leq 
(1-4\zeta^{\scriptscriptstyle
1L}_{3,2}-\gamma\zeta^{\scriptscriptstyle 1L}_{3,2})A_{\tilde l},
\]
provided that $\gamma>1/\zeta^{\scriptscriptstyle 1L}_{3,2}-4$
(this is the usual power-law separating short- from long-range
correlators\cite{halpin-healy1989,balents1993}); for our
correlator with $K_0 \propto 1/u^2$, we indeed obtain that the
tail disappears under the flow and the only acceptable solution is
the exponentially decaying short-range fix-point function realized
for $\zeta = \zeta^{\scriptscriptstyle 1L}_{3,2} \approx
0.174$.

Next, we can make use of the analogy between the dispersive 
four-dimensional
and the non-dispersive three-dimensional case discussed above: 
the fix-point
equations for $R^*$ and for $f^*$ are identical and we obtain an 
exponent
$\sigma= \zeta^{\scriptscriptstyle 1L}_{3,2}\approx 0.174$ 
describing the
power-law type evolution of the correlator's width, cf.\ 
(\ref{ansatz}). An
important difference appears when considering higher-loop 
corrections: such
corrections to $R^*$ and the roughness exponent $\zeta$ are 
relevant for the
$\epsilon$-expansion,\cite{chauve2001,scheidlxxxx} since under the
transformation (\ref{ansatz2}) each higher-loop term appears with 
an additional
factor $\exp[-(\alpha+4\zeta)\, \tilde l] = 1$. On the other 
hand, in the
marginal situation the higher-loop corrections to $f_l$ are 
irrelevant;
higher-loop terms come with an additional factor 
$l^{-(\tau+4\sigma)} = l^{-1}$
and thus can be neglected at large $l$. This happens because at 
the upper critical 
dimension the correlator $K_l$ flows to zero, and higher order 
corrections in 
$K_l$ flow to zero faster and can be neglected, whereas $\tilde 
K_{\tilde l}$ 
(for $\epsilon >0$) flows towards infinity, such that higher 
order terms may be 
important (the growth in amplitude and width turn out to 
compensate each 
other). The above one-loop value for $\sigma$ thus is exact. This 
is similar to 
what happens in critical phenomena, where the leading singularity 
can be 
calculated exactly in one loop at the critical dimension.

We now are in a position to determine the displacement correlator 
$\langle
u^2(\bv{r})\rangle$ on large scales following (\ref{RGeqforu}). 
Taking the
second derivative of (\ref{ansatz}) and using $f_l\to f^*$, we 
obtain $K_{l\geq
l_\mrm{c}}''(0) \sim (16/I) f^{*\prime\prime}(0) 
l^{-(1-2\sigma)}$. We assume
that this asymptotic behavior crosses over to the solution at 
small distances
at $l\sim l_\mrm{c}$. Using $K_{l_\mrm{c}}''(0)=K_0''(0)$, we 
obtain the large
$l$ behavior
\[
   K_{l>l_\mrm{c}}''(0) \sim  K_0''(0)
   \left(\frac{l_\mrm{c}}{l}\right)^{1-2\sigma}.
\]
Integrating (\ref{RGeqforu}), we obtain the displacement 
correlator at large
distances
\begin{eqnarray}
   \langle u^2(\bv{r}) \rangle
   &\sim& I\int_{l_\mrm{c}}^l dl' [-K_{l'}''(0)]\nonumber \\
   &\sim& -IK_0''(0)\int_{l_\mrm{c}}^l dl'
   \left(\frac{l_\mrm{c}}{l'}\right)^{1-2\sigma}\!\!\!\!\!\!\!.
   \label{RGforu2}
\end{eqnarray}
Using (\ref{defxi}) and $l_\mrm{c}=\ln (R_\mrm{c}/\alpha a_0)$, 
we finally
arrive at
\begin{eqnarray}
   \langle u^2(\bv{r})\rangle
   &\sim & \xi^2\left(IK_0^{(4)}(0)l\right)^{2\sigma}
   \label{sublog} \\
   &\sim & \xi^2
   \left( \frac{\ln \left[ \rho(\bv{r})/ a_0\right]}
   {\ln (R_\mrm{c}/\alpha a_0)} \right)^{2\sigma}, \nonumber
\end{eqnarray}
with $\rho$ given by (\ref{rho}). Such a sub-logarithmic behavior 
has already
been conjectured in Ref.\ \onlinecite{blatter1994}; here, we have 
put the
sub-logarithmic growth on a firm basis and have accurately 
calculated the value 
of
the exponent.

The sub-logarithmic growth found in (\ref{sublog}) above has to 
be cut off as
the interaction between the vortices is screened on distances 
beyond the
penetration depth $\lambda$ or when the displacement field 
increases beyond the
lattice constant $a_0$. If the latter condition ($\langle 
u^2(R)\rangle^{1/2}
> a_0$) is realized within the dispersive regime the growth in the
displacement field turns even slower: beyond $u > a_0$ the 
effective disorder
correlator becomes a periodic function in $\bv{u}$, implying that 
$\sigma=0$ in
(\ref{ansatz}) and hence $K_{l>l_\mrm{c}}''(0)\propto 1/l$; the 
integration in
(\ref{RGforu2}) then leads to a growth $\langle u^2(r)\rangle 
\propto \ln \ln
[\rho(\bv{r})/\rho(R_a)]$.\cite{chitra1999} However, typically 
one would expect
that screening becomes relevant before periodicity in $K_0$ and 
the
displacement correlator follows the evolution dictated by the 
usual
non-dispersive random manifold- and Bragg Glass regimes, cf.\
(\ref{randommanifold}) and (\ref{bglog}).

The above results have been derived within a weak collective 
pinning approach
with $R_\mrm{c}>a_0$ and are valid on scales $\rho(\bv{r})> a_0$. 
On the other
hand, increasing the disorder strength, the collective pinning 
radius
$R_\mrm{c}$ drops below $a_0$ and vortices become individually
pinned.\cite{blatter1994} Within this single vortex pinning 
regime, finite
segments of length $L_\mrm{c} \sim 
[\varepsilon_0^2/k_0^{\scriptscriptstyle
(4)}(0)]^{1/3}$ are pinned by the collective action of defects; 
here, $k_0(u)$
denotes the pinning energy correlator for an individual vortex 
line and
$\varepsilon_0 = (\Phi_0/4\pi\lambda)^2$ is the short wave-length 
line tension.
Given the same underlying disorder potential, the single vortex- 
and vortex
lattice correlators are related via $k_0(u) = a_0^2 K_0(u)$. The 
collective
pinning length $L_\mrm{c}$ and the collective pinning radius 
$R_\mrm{c}$ then
are \emph{formally} related via
\begin{equation}
   R_\mrm{c} \sim a_0 \exp{\left[{\rm const.}
   \left(\frac{L_\mrm{c}}{a_0}\right)^3\right]},
   \label{forel}
\end{equation}
cf.\ (\ref{lc}); depending on the disorder strength, either 
$L_\mrm{c} < a_0$
or $R_\mrm{c} > a_0$ is the physically relevant quantity. On the 
other hand,
the length $L_\mrm{c}$ as defined via $k_0^{\scriptscriptstyle 
(4)}(0)$
is a convenient parameter quantifying the strength of the 
disorder.

Assuming $L_\mrm{c} < a_0$ and short distances $L < L_\mrm{c}$ the
mean-squared displacement behaves as
\begin{equation}
   \langle u^2(L)\rangle \sim \xi^2 (L/L_\mrm{c})^3;
   \label{svlarkin}
\end{equation}
the condition $\langle u^2(L_\mrm{c})\rangle \sim \xi^2$ replaces 
the previous
condition $\langle u^2(R_\mrm{c})\rangle \sim \xi^2$ valid in the 
weak pinning
regime. Going beyond the collective pinning (or Larkin-) length 
$L_\mrm{c}$ the
displacement grows as
\begin{equation}
   \langle u^2(L)\rangle
   \sim \xi^2 (L/L_\mrm{c})^{2\zeta^{sv}_{1,2}},
   \label{svrm}
\end{equation}
where $\zeta^{sv}_{1,2}$ is the roughness exponent for the single
vortex.

Let us now connect the single vortex- with the lattice pinning 
regimes at the
scales $L\sim a_0$ and $R\sim a_0$. In the weak pinning case 
($L_\mrm{c}>a_0$), 
we
match (\ref{svlarkin}) with the logarithmic growth 
(\ref{firstlog}) valid for
$a_0<\rho(\bv{r})< R_\mrm{c}/\alpha$,
\[
   \langle u^2(\bv{r}) \rangle
   \sim \xi^2 \left( a_0/L_\mrm{c} \right)^3 \ln
   \left[ \rho(\bv{r})/ a_0\right],
\]
and we recover the relation $\ln [R_\mrm{c}/\alpha a_0] \sim
(L_\mrm{c}/a_0)^3$, in agreement with (\ref{forel}) above. The 
equation for the
sub-logarithmic growth (\ref{sublog}) valid for 
$\rho(\bv{r})>R_\mrm{c}/\alpha$
can then be casted in the form
\[
   \langle u^2(\bv{r}) \rangle
   \sim \xi^2 \left(\left( a_0/L_\mrm{c} \right)^3 \ln
   \left[ \rho(\bv{r})/a_0\right] \right)^{2\sigma}.
\]
For intermediate pinning ($\xi\ll L_\mrm{c}<a_0$) we match
(\ref{svrm}) (valid up to $L \sim a_0$) to the sub-logarithmic
growth (\ref{sublog}) and obtain the relation
\[
   \langle u^2(\bv{r}) \rangle
   \sim \xi^2 \left( a_0/L_\mrm{c} \right)^{2\zeta^{sv}_{1,2}}
   \left( \ln \left[ \rho(\bv{r})/a_0\right]\right)^{2\sigma}
\]
describing the growth of the displacement correlator for larger 
distances
$\rho(\bv{r})>a_0$.

Comparing the two cases of weak and intermediate pinning in the 
sub-logarithmic
regime we note a slightly different dependence on the disorder 
parameter
$L_\mrm{c}$: in the weak pinning situation the disorder 
dependence is given by
the factor $L_\mrm{c}^{-6\sigma}$, whereas for intermediate 
pinning the
corresponding factor $L_\mrm{c}^{-2\zeta^{sv}_{1,2}}$ involves a 
different
exponent. The numerical values of the two exponents can be 
calculated within
RG; the individual vortex is characterized by an isotropic 
elasticity and
$\zeta^{sv}_{d,2}\approx 0.177\epsilon$ to first loop 
order.\cite{emig1999}
This value is close to the one obtained for the anisotropic 
elasticity of the
vortex lattice, $\zeta_{d,2}\approx 0.174\epsilon$. Therefore, 
the exponent
$2\zeta^{sv}_{1,2} \approx 1.062$ is similar to the value $6 
\sigma \approx
1.044$. However, note that the roughness exponent 
$\zeta^{sv}_{1,2}$ is subject
to corrections by higher-loop terms, whereas the value for 
$\sigma$ remains
unchanged.

\section{Conclusion}

Summarizing, the long-ranged interaction between vortices as 
expressed by the
dispersive elastic moduli of the flux-line lattice strongly 
reduces the effect
of disorder: rather than the usual algebraic growth $\langle 
u^2(\bv{r})
\rangle \propto r^{2\zeta}$ of the mean-squared displacement 
correlator, one
finds a mere logarithmic dependence in the perturbative regime 
$a_0<\rho(\bv{r})
< R_\mrm{c}/\alpha$,\cite{larkin1979}
\[
   \langle u^2(\bv{r}) \rangle
   = \xi^2 \frac{\ln \left[\rho(\bv{r})/ a_0\right]}
   {\ln \left[R_\mrm{c}/ \alpha a_0\right]},
\]
with $\rho(\bv{r})$ given by (\ref{rbar}). On larger scales 
$R_\mrm{c}/\alpha <
\rho(\bv{r}) < \lambda$ we have found a sub-logarithmic law,
\[
   \langle u^2(\bv{r}) \rangle
   \sim \xi^2 \left( \frac{\ln \left[ \rho(\bv{r})/a_0\right]}
   {\ln (R_\mrm{c}/\alpha a_0)}\right)^{2\sigma};
\]
here, the scale function $\rho(\bv{r})$ is determined to the 
precision
given in (\ref{rho}) with $\alpha= {\alpha'}^2\approx 0.32$.
The exponent $2\sigma \approx 0.348$ obtained within the one-loop 
RG calculation
remains unchanged by higher loop corrections. The definition
\begin{equation}
\xi^2 = \frac{-K_0''(0)}{K_0^{(4)}(0)}
\end{equation}
for the relevant length scale of the disorder landscape emerges 
naturally
within the RG framework.

The mean-squared displacement correlator for the disordered 
vortex lattice then
exhibits two separate regimes with a well developed translational 
order
involving either a logarithmic or a sub-logarithmic growth of 
$\langle
u^2(\bv{r})\rangle$, a first regime at small length scales 
originating from the
long-ranged interaction between vortices, and a second regime at 
large scales
which is due to the periodicity of the lattice (Bragg Glass 
regime, cf.\ Eq.\
(\ref{bglog})). A number of experiments have been devised in 
order to observe
this type of logarithmic order, e.g., via Bitter
decoration\cite{murray1990,kim1999} or using small angle neutron 
scattering
(SANS)\cite{gammel1998,klein2001} (we note that the Bitter 
decoration analysis
carried out at low fields does not probe the dispersive regime 
discussed here).
The observation of a well (i.e., quasi-long
range) ordered lattice then is often interpreted as evidence for 
Bragg-glass
order. The above discussion demonstrates, that this need not to 
be the case,
however, as the vortex lattice exhibits this type of logarithmic 
order on
intermediate scales $a_0 < R < \lambda$, $a_0 < L < 
\lambda^2/a_0$ as well
(note the `cigar' shaped geometry of the dispersive regime which 
can reach a
large extension along the longitudinal direction, i.e., parallel 
to the
magnetic field).

A further complication of this type of analysis is found in the 
slow relaxation
of the vortex system towards the proper glassy order: frozen-in 
at some higher
temperature $T_{\rm melt}$ where pinning is irrelevant, the 
glassy order of the
vortex lattice first has to establish itself via proper 
relaxation through
creep over the relevant barriers. During the experimental time 
$t$, the vortex
system then can overcome barriers of size $T\ln(t/t_0)$ 
typically, with
$t_0^{-1}$ a proper attempt frequency\cite{blatter1994}. With 
typical
temperatures of order 50 K and $\ln(t/t_0) \approx 20$, barriers 
$U(R) \approx
1000$ K may be overcome, producing relaxed domains of size ($R$) 
of a few
lattice constants typically; hence reaching glassy order over 
large scales is
quite a nontrivial task.

\begin{acknowledgments}

The authors wish to thank Konstantin Efetov, Denis Gorokhov,
Anatoly Larkin, and Valerii Vinokur for helpful discussions.
Financial support from the Swiss National Foundation is gratefully
acknowledged.

\end{acknowledgments}

\bibliographystyle{apsrev}

\end{document}